\documentclass[conference]{IEEEtran}
\IEEEoverridecommandlockouts
\usepackage{cite}
\usepackage{amsmath,amssymb,amsfonts}
\usepackage{algorithmic}
\usepackage{graphicx}
\usepackage{textcomp}
\usepackage{xcolor}
\usepackage{url}
\usepackage[colorlinks]{hyperref}
\hypersetup{
     colorlinks   = true,
     linkcolor    = blue,
     urlcolor     = blue,
     citecolor    = blue
}
\usepackage{comment}

\usepackage{flushend}
\def\BibTeX{{\rm B\kern-.05em{\sc i\kern-.025em b}\kern-.08em
    T\kern-.1667em\lower.7ex\hbox{E}\kern-.125emX}}
\begin{document}

\title{
MANA-2.0:  A Future-Proof Design for \\
Transparent Checkpointing of MPI at Scale


{\thanks{\noindent $^*$This work was supported by the Office of
Advanced Scientific Computing Research in the Department
of Energy Office of Science under contract number DE-AC02-
05CH11231. \\
\hbox{~~} $^\dag$This work was partially supported by National
Science Foundation Grant OAC-1740218 and a grant from
Intel Corporation.}
}
}

\author{\IEEEauthorblockN{Yao Xu$^\dag$}
\IEEEauthorblockA{\textit{Khoury College of Computer Sciences} \\
\textit{Northeastern University}\\
Boston, USA \\
xu.yao1@northeastern.edu}
\and
\IEEEauthorblockN{Zhengji Zhao$^*$}
\IEEEauthorblockA{\textit{NERSC} \\
\textit{Lawrence Berkeley National Laboratory}\\
Berkeley, USA \\
zzhao@lbl.gov}
\and
\IEEEauthorblockN{Rohan Garg}
\IEEEauthorblockA{
\textit{  ~~~~~~~~~~~~~~~~~ Nutanix, Inc. ~~~~~~~~~~~~~~~~~ }\\
Seattle, USA \\
rohan.garg@nutanix.com}
\and
\IEEEauthorblockN{Harsh Khetawat}
\IEEEauthorblockA{\textit{Department of Computer Science} \\
\textit{North Carolina State University}\\
Raleigh, USA \\
hkhetaw@ncsu.edu}
\and
\IEEEauthorblockN{Rebecca Hartman--Baker$^*$}
\IEEEauthorblockA{\textit{NERSC} \\
\textit{Lawrence Berkeley National Laboratory}\\
Berkeley, USA \\
rjhartmanbaker@lbl.gov}
\and
\IEEEauthorblockN{Gene Cooperman$^\dag$}
\IEEEauthorblockA{\textit{Khoury College of Computer Sciences} \\
\textit{Northeastern University}\\
Boston, USA \\
gene@ccs.neu.edu}
}
\maketitle

\begin{abstract}
MANA-2.0 is a scalable, future-proof design for transparent checkpointing of MPI-based computations. Its network transparency (``network-agnostic'') feature ensures that MANA-2.0 will provide a viable, efficient mechanism for transparently checkpointing MPI applications on current and future supercomputers. MANA-2.0 is an enhancement of previous work, the original MANA, which interposes MPI calls, and is a work in progress intended for production deployment. MANA-2.0 implements a series of new algorithms and features that improve MANA’s scalability and reliability, enabling transparent checkpoint-restart over thousands of MPI processes. 
MANA-2.0 is being tested on today's Cori supercomputer at NERSC using Cray MPICH library over the Cray GNI network, but it is designed to work over any standard MPI running over an arbitrary network.
Two widely-used HPC applications were selected to demonstrate the enhanced features of MANA-2.0: GROMACS, a molecular dynamics simulation code with frequent point-to-point communication, and VASP, a materials science code with frequent MPI collective communication.
Perhaps the most important lesson to be learned from MANA-2.0 is a series of algorithms and data structures for library-based transformations that enable MPI-based computations over MANA-2.0 to reliably survive the checkpoint-restart transition. 
\end{abstract}


\section{Introduction}

MANA (MPI-Agnostic Network-Agnostic transparent  checkpointing tool) is a previously developed package for checkpointing MPI applications~\cite{garg2019mana}.  Among its unique features, MANA supports checkpoint-restart for MPI applications, while being transparent to (a)~the MPI application; (b)~the MPI library itself; and (c)~the network libraries underlying the MPI library.  These features are based on a novel split-process architecture (see~\ref{sec:splitProcess}).  Unlike previous approaches, MANA directly and transparently interposes the MPI calls themselves, taking advantage of the standardized API for MPI.

Transparent checkpointing is a prerequisite for system-level checkpointing, a vital tool for the operational needs of computing centers. System-level checkpointing is used to manage jobs in a real-world environment, which includes outage and maintenance events, and workloads with differing priorities and turnaround-time expectations. For example, experimental and  observational facilities supported by the U.S\hbox{.} Department of Energy's Office of Science (DOE-SC) often require high-priority, real-time access to computing resources to  generate  immediate  feedback  for  follow-up  experiments.



While many scientific applications employ some level of internal checkpoint-restart (C/R) support, they lack a standard API for checkpoint and restart. And they usually require waiting for a particular computation phase (e.g., after an iteration completes).  For example, when chaining together allocation slots for a long-running execution, the inability to guarantee a checkpoint within the last half hour of an allocation makes its use inflexible~\cite{cr-vision2021}. 

In addition, application-level checkpoint-restart may support some application features and not others, especially for applications with a large code base.  As an example, VASP~\cite{vasp} has internal C/R support for atomic relaxation and MD simulations, but not for Random Phase Approximations.
VASP has tens of thousands of users, and it consumes about 20\% of computing cycles at the NERSC supercomputing center\footnote{NERSC (National Energy Research Scientific Computing) is the primary computing facility for the U.S. Department of Energy's Office of Science (\url{https://www.nersc.gov/}).}.  Transparent checkpointing helps VASP users to chain long-running jobs, while allowing NERSC to shift this 20\% of its computing resources upon arrival of a large, real-time workloads as described above.

MANA-2.0 is intended as a scalable, future-proof design for transparent checkpointing of MPI-based computations.  While MANA is being tested on NERSC's Cori supercomputer using the Cray GNI network, its network transparency (``network-agnostic'') ensures that MANA will provide a viable, efficient mechanism for transparently checkpointing computational workloads on both current and future supercomputers.




MANA's unique feature of MPI and network transparency is a good fit for NERSC,
which employs supercomputing resources with a proprietary Cray GNI network.  The development of MANA is the basis for a long-term collaboration between NERSC and the MANA team.  Note that previous approaches to checkpointing MPI~\cite{bouteiller2006mpich,ansel2009dmtcp,gao2006application,hursey2007design,cao2014transparent} supported TCP and InfiniBand, but did not support the Cray GNI network.  The original MANA was the first package that could support Cray GNI, through a new split-process architecture~\cite{garg2019mana}.

This initial promise led to a goal of supporting production-level deployment of MANA on NERSC's Cori supercomputer~\cite{zhao2020deploying}; and following that, to support Perlmutter (the \#5 supercomputer in the world as of this writing~\cite{top500jun2021}).
But history has shown achieving the goal of production-level transparent checkpointing to be more difficult than anticipated.
In personal communications, the authors of the original MANA~\cite{garg2019mana} and of an updated version~\cite{chouhan2021improving} have both expressed concerns about the fragility of the software architecture and its key components.

In general, the challenge of developing robust checkpointing algorithms for MPI can be understood by analogy to developing a new optimizing compiler.
Both endeavors are concerned with translating high-level code representations into lower-level executions. Both endeavors require subtle algorithms to preserve the high-level semantics when translated to a lower level (compilers), or across the checkpoint-restart boundary (MANA).

This work demonstrates a robust version of MANA (MANA-2.0) that can scale to a high level.  This work provides two novel elements:
\begin{enumerate}
     \item While still a work in progress, MANA-2.0 is already shown to scale well on two very different types of computations (see Section~\ref{sec:experiment}).  GROMACS~\cite{gromacs} highlights intensive MPI point-to-point communication.  VASP~\cite{vasp} highlights intensive MPI collective communication.  VASP is responsible for more than 20\% of the machine time on Cori~\cite{zhao2020deploying}.
     \item Perhaps even more important, MANA-2.0 is the fruit of a series of enhancements that serve as lessons learned for related research projects.  MANA-2.0 applies a wrapper-function based strategy to \emph{efficiently} translate each MPI call of the original MPI application into one or more direct MPI calls.  The algorithms of MANA-2.0 include data structures that enable MANA-based computations to survive a checkpoint and full restart.  In this respect, MANA-2.0 bears as much resemblance to a compiler (translating into lower level MPI calls), as it does to a simple library utility.
\end{enumerate}

The two novel elements mentioned above represent a qualitative difference of the current work over the original MANA.  For the first time, MANA-2.0 demonstrates the ability to \emph{reliably} checkpoint GROMACS, even at 2048 MPI processes.
In comparison, the original MANA work~\cite{garg2019mana} was intended as a proof-of-concept, thus it was not able to reliably checkpoint-restart at this scale. 


MANA-2.0 is an example of a larger class of projects that rely on source-level transformations of MPI calls while maintaining correctness and performance.  See the beginning of Section~\ref{sec:componentAlgorithms} for a list of issues in MPI source transformations that were found to be important for correctness and performance.  A short list of the relevant issues (expanded on in Section~\ref{sec:componentAlgorithms}) includes:
(i)~decomposition of blocking MPI calls into asynchronous calls (e.g., \texttt{MPI\_Send} to \texttt{MPI\_Isend/MPI\_Test});
(ii)~insertion of blocking MPI calls while avoiding deadlock (e.g., is inserting \texttt{MPI\_Barrier} before \texttt{MPI\_Bcast} valid?);
(iii)~determining whether an MPI call can be satisfied locally (e.g., \texttt{MPI\_Translate\_group\_ranks}); and
(iv)~when can it be proved that a (virtualized) MPI request object can no longer be accessed by the MPI application in the future.

This work is organized into the following sections.
Section~\ref{sec:background} briefly describes the underlying split-process design of the original MANA, then provides further details of how checkpointing is supported for key components of MPI.
Section~\ref{sec:componentAlgorithms} describes the algorithmic innovations of this work in fixing many of the deficiencies in key components of MANA.
Section~\ref{sec:experiment} presents an experimental evaluation of the modified version of MANA.
Section~\ref{sec:relatedWork} presents related work.
Finally, Section~\ref{sec:conclusion} is the conclusion.

\section{Background}
\label{sec:background}



\subsection{Split-processes}
\label{sec:splitProcess}

In brief, the key idea of a \emph{split-process} approach is to load two independent programs into the virtual memory of a single process.  Because they are contained within the same virtual memory, a function from one program (typically the ``upper-half'' program) may call a function of the other program (typically the ``lower-half'' program) --- so long as the address of the lower-half function is known to the upper-half function.

\begin{figure}[!tb]
\centering
\includegraphics[width=1\columnwidth]{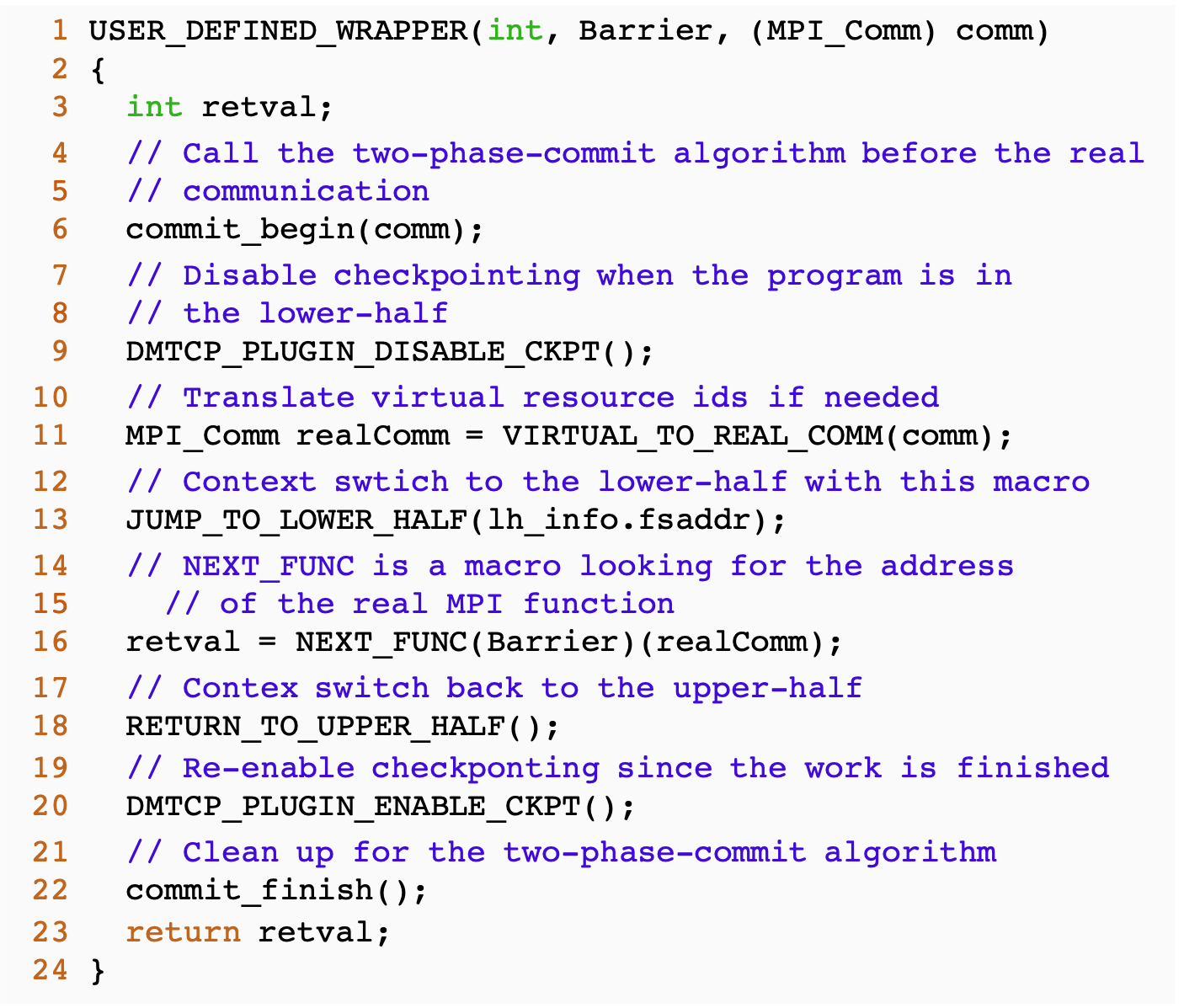}

\caption{Code snippet of the MPI\_Barrier wrapper.}
\label{fig:wrapper_snippet}
\end{figure}

In practice, the upper-half program will be the MPI application program, dynamically linked with a ``stub'' MPI library.  The ``stub'' MPI library consists of wrapper functions around each MPI call.  The wrapper calls a lower-half function in the actual MPI library.  Finally, the lower-half program consists of a small MPI application linked to the actual MPI library, which links to the necessary libraries. Figure~\ref{fig:wrapper_snippet} shows an example of the \texttt{MPI\_Barrier} wrapper. 

The advantage of this scheme is that only the upper-half program is checkpointed.  (Only its memory is saved in a checkpoint image file.)  This sidesteps the key issue of other checkpointing approaches:  There is no need to disconnect and re-connect the network (the Cray GNI network in our case).
At the time of restart, the lower-half program is started, and it loads the upper-half program into memory at the original address, from the checkpoint image file. 
For a deeper description of split processes, see the original paper of Garg \hbox{et al.}~\cite{garg2019mana}.

\subsection{Overview of Semantic Components of MANA}
\label{sec:semantics}

We standardize here on MPI-3.1~\cite{mpi-standard-3.1}.  There are primarily four categories for which MANA must save state at the time of checkpoint:
\begin{enumerate}
    \item the state of all memory in the upper half;
    \item a consistent snapshot associated with any MPI collective communication calls in progress (e.g., \texttt{MPI\_Barrier} or \texttt{MPI\_Bcast};
    \item a consistent snapshot associated with any MPI point-to-point calls in progress (e.g., \texttt{MPI\_Send} and \texttt{MPI\_Recv};
    \item any MPI one-sided communication calls (the \texttt{MPI\_Win\_XXX} family of calls).
\end{enumerate}

MPI's one-sided communication calls are not yet supported, but support for the \texttt{MPI\_Win\_} family is on the roadmap of MANA.  
Details on the remaining three categories are described in the next section.

\subsection{Virtualized MPI objects}

MPI calls may create new objects of types such as \texttt{MPI\_Comm} and \texttt{MPI\_Request}.  In the MANA wrapper functions around these calls, a new virtual object (virtualized communicator or virtualized request in our example) is created and returned to the user's MPI application.  An internal mapping from the virtual object to the ``real'' object returned by the lower-half MPI library is maintained.  Thus, when the user's MPI application makes a later call, using one of these virtualized objects, the MANA wrapper function automatically replaces the virtualized object by the real object stored in its mapping.

This is important, since an MPI application may make copies of its objects, to be stored at arbitrary addresses.  At restart time, MANA simply updates its virtual-to-real mapping with new, real objects, instead of trying to directly patch the memory of the MPI application with updated real objects.

\section{Novel algorithms for key components of MANA}
\label{sec:componentAlgorithms}


We begin this section by enumerating the challenges faced in bringing the original MANA proof-of-principle closer to a production standard, capable of supporting C/R in production. The subsections following this discussion describe individual, novel algorithms that were introduced with MANA-2.0 to improve its reliability and performance.

The challenges in supporting MANA robustly can be attributed to several factors.
\begin{enumerate}
    \item MANA is unusual in interposing directly at the level of the MPI API.  A checkpoint can be taken \emph{only} if no MPI process is in the middle of the MPI library.  Hence, MANA uses wrapper functions to interpose between some MPI calls to \texttt{MPI\_Send} and \texttt{MPI\_Recv}, and then redirects to asynchronous calls like \texttt{MPI\_Isend} and \texttt{MPI\_Irecv}.  As another example, MANA  interposes calls to \texttt{MPI\_Wait} and redirects to a loop around \texttt{MPI\_Test}.  These asynchronous calls are non-blocking, and so MANA can guarantee not to checkpoint a send/receive in the middle of a checkpoint.  Extensions to \texttt{MPI\_THREAD\_MULTIPLE} are not considered here.  Subsections~\ref{sec:A-virtReq} and~\ref{sec:B-drainSendRecv} discuss how to handle the conversion of MPI point-to-point communications correctly.
    \item The conversion to semantically equivalent MPI calls can result in higher runtime overhead. Subsection~\ref{sec:D-collectivePerformance} shows an example of this type of runtime overhead.
    Nevertheless, semantic conversions are required for multiple reasons.  The previous item gave an example, transforming \texttt{MPI\_Send}/\texttt{MPI\_Recv} to \texttt{MPI\_Isend}/\texttt{MPI\_Irecv}.  In a similar example, \texttt{MPI\_Wait} is converted to multiple calls to \texttt{MPI\_Test}.  And in some cases, an MPI call is converted to a POSIX system call, as in the conversion of \texttt{MPI\_Alloc\_mem}/\texttt{MPI\_Free\_mem} to malloc/free.
    \item A conversion to semantically equivalent MPI calls, while valid for most MPI implementations, cannot be guaranteed for all MPI implementations.  In particular, see~\cite{mpi-semantics}, the MPI-4.0 addendum for semantics.  This helps resolve questions such as: when it is valid to add an \texttt{MPI\_Barrier} in front of a \emph{non-blocking} MPI collective communication (e.g., the root in \texttt{MPI\_Bcast}); whether the insertion of an \texttt{MPI\_Barrier} will slow down or accelerate an MPI application (see~\cite[page~41: MPICH\_COLL\_SYNC]{cray-mpi}); and which MPI calls may be resolved solely using local information.  \texttt{MPI\_Barrier} before \texttt{MPI\_Bcast} also can create deadlock. Details are discussed in Section~\ref{sec:E-collectiveCorrectness} of how \texttt{MPI\_Bcast} is supported, while avoiding potential deadlock.
    \item While MANA can use its centralized coordinator as a side channel to communicate among the processes, this is inefficient.  Hence, MANA-2.0 instead makes direct use of MPI calls as being more maintainable and more efficient, while making sure semantically to be non-intrusive. Subsections~\ref{sec:B-drainSendRecv} and~\ref{sec:K-globallyUniqueID} are examples of this practice.
    
    For example, the original MANA needed to ``drain'' any point-to-point messages (\texttt{MPI\_Send} and \texttt{MPI\_Recv}).  Previously, the DMTCP coordinator tracked the total number of bytes sent and received.  In MANA-2.0, at the time of checkpoint,
    \texttt{MPI\_Alltoall} is used to directly communicate among the MPI processes to determine if the number of bytes sent equals the number of bytes received (implying that there are no pending bytes in the network).  While Subsection~\ref{sec:B-drainSendRecv} discusses draining messages, this last point is covered at the end of Subsection~\ref{sec:M-lessonsLearned}.
    \item As part of the draining of point-to-point messages discussed in the last item, all MPI\_Processes need a way to unambiguously identify a unique MPI process.  This can be done by using \texttt{MPI\_Translate\_group\_ranks} to translate the MPI rank within the current communicator to the MPI rank within the world communicator.
    MANA-2.0 takes care to internally use MPI calls that solely access local information, where possible, for the sake of efficiency.  However, the overhead can be still sensitive to the particular MPI implementation.  An implementation may choose to implement these MPI calls as ``local'' calls, or else it may choose a simpler implementation that invokes communication with a central MPI resource manager.  Subsections~\ref{sec:B-drainSendRecv} and~\ref{sec:K-globallyUniqueID} show how MPI calls are used internally in MANA-2.0.
    \item MANA-2.0 improves on the original implementation that virtualizes MPI objects.  For example, the original MANA virtualizes MPI communicators, and MANA-2.0 additionally virtualizes MPI requests.  Virtualization of objects or IDs is a technique from \emph{process virtualization}~\cite{arya2016design}.  A given communicator, request or other object is associated both with a real ID (known within the MPI library) and with a virtual ID (stored within the MPI application memory).  MANA does the translation when it interposes on any calls from the MPI application to the MPI library.
    Thus, if a checkpoint-restart occurs between the creation of the object and a second use, then the virtualized object can later be bound to a newly created real object on restart.
    
    However, when a real object is no longer referenced within the MPI library, this may be unknown to MANA, and MANA can be left with a growing list of stale virtualized objects. The size of that list continues to grow if it cannot be garbage-collected --- resulting both in a growing memory footprint, and in higher overhead to access an object. Subsection~\ref{sec:A-virtReq} discusses details of virtualizing MPI requests.
    \item The original MANA has a large runtime overhead, especially when used with applications with intensive collective communications. Subsections~\ref{sec:G-fs-register}, \ref{sec:H-lambda-functions}, \ref{sec:I-runtimeOverhead}, \ref{sec:J-stragglers}, \ref{sec:K-globallyUniqueID} and~\ref{sec:L-hybrid} focus on addressing the performance issues of the original MANA.
    
    \item The original MANA supports Fortran MPI programs with a series of wrappers that translate MPI's Fortran bindings to MPI's C bindings. However, some corner cases are not handled correctly due to differences between the Fortran and C languages. \ref{sec:F-fortran} shows one example of such differences.


\end{enumerate}

\bigskip
The following subsections discuss these and other issues that arose.

\subsection{Virtualized requests}
\label{sec:A-virtReq}

Resources like \texttt{MPI\_Comm} and \texttt{MPI\_Group} are allocated by MPI libraries.
The resources must be virtualized to survive the checkpoint-restart barrier.  But the \texttt{MPI\_Request} resource was not virtualized in the original MANA.
This issue was not seen in the original implementation~\cite{garg2019mana} because non-blocking collective communications were not supported.

Compared to other virtualized resources, virtual MPI requests are generated so frequently that one must aggressively prune completed virtual MPI requests to avoid large performance and memory overhead. New virtual MPI requests are created in non-blocking MPI function wrappers, and retired in \texttt{MPI\_Test} and \texttt{MPI\_Wait} wrappers.  Recall that the MPI library sets a request to \texttt{MPI\_REQUEST\_NULL} when the request has been satisfied.
Similarly in MANA-2.0, when MANA wishes to ``retire'' or garbage-collect the virtualized form of an MPI request, it deallocates its request from its internal table, and sets the MPI request value in application memory to \texttt{MPI\_REQUEST\_NULL}.

An alternative implementation was suggested by one reviewer and is to be investigated.  MANA-2.0 could internally use \texttt{MPI\_Request\_get\_status} to non-destructively interrogate the MPI library.  As a result, the application's MPI request status will be in a known (non-null) state, and so it will be safe to later internally call \texttt{MPI\_Test} instead of directly setting an MPI request value to \texttt{MPI\_REQUEST\_NULL} in application memory.

MPI requests are used in two different cases:  non-blocking collective communication and point-to-point communication. Because these two cases require different algorithm and data structures to support checkpoint/restart, virtualized MPI request retirements need to be supported differently for each case.

In the case of non-blocking collective communication, a log-and-replay algorithm is used to support checkpoint/restart. Upon a successful \texttt{MPI\_Test} or \texttt{MPI\_Wait}, virtualized MPI requests can be removed immediately from MANA's internal table without interfering with replaying on restart. Since the addresses of the tested MPI request is known in wrapper functions of \texttt{MPI\_Test} and \texttt{MPI\_Wait}, the MPI requests values in application's memory can simply be set to \texttt{MPI\_REQUEST\_NULL} directly.

The second use of MPI requests is to handle non-blocking point-to-point communication.  In this case, values of virtual requests are saved in a list of active non-blocking calls, and addresses of MPI requests in the application's memory are unknown.  Therefore, we cannot update the user application's memory to update the request to \texttt{MPI\_REQUEST\_NULL} directly. 

Since we cannot directly update the user's memory in this case, a two-step retirement algorithm was developed to safely delete completed requests without requiring knowledge of the addresses where the user application may have stored the request.  When a request is complete, we update the virtual ID table so that the completed virtual request points to a special value \texttt{MPI\_REQUEST\_NULL}. The next time \texttt{MPI\_Test} and \texttt{MPI\_Wait} are called, we know the virtual request is ready to be removed, since the real request is \texttt{MPI\_REQUEST\_NULL}. We can then safely remove the virtual request from the table and set the user application's request variable to \texttt{MPI\_REQUEST\_NULL}.

\subsection{Drain send-receive for point-to-point communication}
\label{sec:B-drainSendRecv}

In the previous work~\cite{garg2019mana}, MANA translated blocking point-to-point communications to comparable non-blocking versions, and used a variation of an all-to-all exchange algorithm to perform bookkeeping when draining point-to-point messages in the network during checkpoint. Point-to-point communication wrappers accumulated the count of the number of messages at runtime. When checkpointing, each process sent the counted number of messages to the coordinator, and the coordinator sent the total number of messages to each process. If the total send and receive counts did not match, MANA used \texttt{MPI\_Iprobe} to detect messages still in the network and tried to receive them with \texttt{MPI\_Recv}. Finally, MANA updated the new send and receive counts to the coordinator and repeated the process.

This design has some drawbacks, however: Frequent communication with the coordinator can be expensive when running at large scale, and sharing only the total number of sends and receives makes it impossible to do any debugging to determine to which MPI rank a missing message might belong. 

Therefore, MANA-2.0 improves the algorithm by using a smaller-grain message counter for each pair of process and sharing only essential information with \texttt{MPI\_Alltoall}. After calling \texttt{MPI\_Alltoall} at checkpoint time, all processes know, without further communication, how many bytes they were expected to receive and how many bytes they actually received. Locally, each process is able to use \texttt{MPI\_Recv} to drain missing bytes from peers.

Another issue addressed in MANA-2.0 is that if \texttt{MPI\_Recv} or \texttt{MPI\_Irecv} has already been called, then the message may have already been received, in which case \texttt{MPI\_Iprobe} can no longer detect the message in the network. Therefore, if some process found no messages in the network by using \texttt{MPI\_Iprobe}, and if the send-receive count is still unbalanced, then there must be an unfinished \texttt{MPI\_Irecv} for which the corresponding request has not yet been satisfied. In this case, instead of using an extra \texttt{MPI\_Recv} to drain the message, we call  \texttt{MPI\_Test} on existing \texttt{MPI\_Irecv} records to discover pending MPI requests associated with \texttt{MPI\_Irecv} records, and to then drain the missing messages.

\subsection{Keep a list of only the \emph{active} communicators for the sake of restart}
\label{sec:C-activeCommunicators}

In the original design, when restoring MPI communicators during restart, all functions used to create communicators were recorded and replayed. Therefore, many communicators no longer being used would be recreated during restart. In addition, MPI communicators could not be retired, in case they had been used to create other communicators. As a result, time was wasted on replaying unnecessary functions. The virtual ID table (mapping) for communicators also occupied more memory and slowed down the lookup performance.

The new design in MANA-2.0 instead keeps a list of active communicators and groups, and reconstructs only communicators and groups in the active list during restart.  A knowledge of the underlying MPI group and its members suffices to recreate a semantically identical communicator.  So, it is no longer necessary to replay MPI calls that build new communicators from old communicators.


\subsection{Not all collective communications are barriers:  performance issues}
\label{sec:D-collectivePerformance}

In the MPI standard~\cite{mpi-standard-3.1}, there is no requirement in a collective function that all participating MPI processes must enter the function before any process can return. (For example, the ``root'' in \texttt{MPI\_Bcast} can broadcast its message and return before other processes receive the message.) However, because of the two-phase-commit algorithm of the original MANA (see~\cite{garg2019mana}), a barrier was added before each collective communication call.  This was done to avoid checkpointing in the middle of collective calls, such as \texttt{MPI\_Bcast}.  However, this changes the semantics by making \texttt{MPI\_Bcast} a blocking call.  (See the next subsection for how MANA-2.0 restores the semantics of \texttt{MPI\_Bcast} as a non-blocking call.)

The modified semantics incurs a major performance impact for collective communications. For example, adding a barrier before an \texttt{MPI\_Bcast} forces the ``root'' process to wait until all other processes arrives. Generally the barrier makes the \texttt{MPI\_Bcast} run two to three times slower. (See the next subsection for the solution in MANA-2.0.)  Nevertheless, in the case of \texttt{MPI\_Allreduce}, where all processes need to send and receive data from other processes, the barrier slightly improved the performance in our tests. Hence, note the recommendation of Cray for optionally adding a barrier to end-user code, and then testing with CRAYPAT to see if the performance improves~\cite[page~41]{cray-mpi}.

\subsection{Not all collective communications are barriers:  correctness issues}
\label{sec:E-collectiveCorrectness}

In addition to the impact on performance, in some rare cases the added barrier can lead to a deadlock that did not exist in the native MPI application. Assume a scenario where two MPI processes (rank~0 and rank~1) communicate with each other. Let rank~0 call \texttt{MPI\_Bcast} as the ``root'' rank.  In this scenario, rank~0 first calls \texttt{MPI\_Send} and then calls \texttt{MPI\_Bcast}.  And rank~1 calls \texttt{MPI\_Recv} and then calls \texttt{MPI\_Bcast}. There is no deadlock when running natively, since rank~0 (the root rank) will not block on \texttt{MPI\_Bcast}. However, if MANA adds a barrier before the \texttt{MPI\_Bcast}, then rank~0 will wait for rank~1 to join the barrier, and rank~1 can only join the barrier after receiving in \texttt{MPI\_Recv}.  Yet, rank~0 has not yet called \texttt{MPI\_Send}. Therefore, we have a deadlock.

This deadlock occurred in the original MANA because it converted the semantics of \texttt{MPI\_Bcast} to a blocking call.
To avoid the deadlock, MANA-2.0 must restore the non-blocking (but synchronizing) semantics of the MPI standard (see MPI-4, semantics appendix~\cite{mpi-semantics}).  So, MANA-2.0 provides alternative wrapper implementations for \texttt{MPI\_Bcast} and similar MPI calls, which use point-to-point communication (\texttt{MPI\_Send} and \texttt{MPI\_Recv}) instead of the real \texttt{MPI\_Bcast} function in the lower half.  (Note that the new persistent collectives that are part of the MPI-4 standard are outside the scope of this discussion.)

Normally, this will cause a performance degradation. Typically, the use of point-to-point communication compared to the standard \texttt{MPI\_Bcast} implementation. However, a new hybrid two-phase-commit algorithm is currently being developed. The new algorithm only inserts the trivial barrier at checkpoint time. Similarly, the MANA wrapper function for \texttt{MPI\_Bcast} in the new algorithm will use the standard \texttt{MPI\_Bcast} before checkpoint. At checkpoint time, all calls to \texttt{MPI\_Bcast} will use the alternative point-to-point implementation.



\subsection{Handling MPI named constants in Fortran}
\label{sec:F-fortran}

Because of the nature of Fortran, some MPI named constants, such as \texttt{MPI\_IN\_PLACE} and \texttt{MPI\_STATUS\_IGNORE}, are set at link time instead of compile time~\cite{zhang2014implementing}. This is because Fortran uses common blocks, instead of true global constants.  So, named constants in Fortran are addresses of unique storage locations in the underlying MPI library. Therefore, when using MANA with Fortran-based MPI applications, the named constants passed into MANA's Fortran wrappers are addresses, instead of the actual constant values as in the C~interface.  See~\cite{zhang2014implementing} for details.
(Note that Fortran~2018 has recently introduced a more direct way for Fortran and the C~language to communicate Fortran constants.)

To identify these link-time named constants correctly in MANA's wrappers, we linked a small Fortran routine into MANA that discovers the value/address of the Fortran named constants dynamically. If a parameter passed in from a user's application matches a Fortran named constant, then MANA-2.0 replaces the value with the equivalent C constant when calling the real MPI function in the lower half.

\subsection{The FS register}
\label{sec:G-fs-register}

A major source of runtime overhead is due to the use of the ``FS'' register.  The split-process model of MANA~\cite{garg2019mana} requires an upper-half program to do a context switch to a lower-half program.  This requires MANA to modify the FS register.  Unfortunately, that operation has been inordinately expensive (microseconds or more) due to the need to make a kernel call to the Linux kernel.  Only recently, the unprivileged use of the FS register was enabled in Linux~5.9~\cite{linux-fsgsbase}.  Most HPC sites conservatively use older Linux kernels.
For systems that cannot use the FSGSBASE Linux kernel patch, MANA-2.0 added a workaround to reduce the cost of using the "FS" register.  For details, see~\cite{yao2021removing}.

\subsection{C++ lambda functions}
\label{sec:H-lambda-functions}

C++ lambda functions were used in many MPI function wrappers in MANA to increase the readability of codes, but they come at the cost of performance.  The C++ compiler had compiled the lambda function of the source code into three or four additional call frames in the binary.  For frequent MPI calls, this can add significant runtime overhead. To remove lambda functions in MANA, functions that take lambda functions as callbacks are decomposed into dedicated ``prepare'' and ``finish'' functions. 


\subsection{Other sources of runtime overhead}
\label{sec:I-runtimeOverhead}

Additional sources of runtime overhead are described in the public documentation of internals provided by MANA-2.0~\cite{mana-documentation}.
These smaller factors also contribute to MANA's runtime overhead.  A brief list of these factors follows.

\begin{enumerate}
    \item Translating virtual ID to real ID depends on map operations of C++ \texttt{std::map}. Typically C++ \texttt{std::map} requires $O(\log{}n)$ to look up an entry in the map. In some cases, the original MANA also uses a linear search in the map.  This can be reduced by employing a C++ map based on hash arrays.
    \item Calls to disable and enable DMTCP checkpoint are used widely in MPI function wrappers. The cost of lock operations are too expensive because of their high frequency of use.
    \item An internal helper method that translates the local rank of a communicator to a global rank makes multiple calls to the lower half.  Calls to the lower half adjust the FS register, which is expensive (see Subsection~\ref{sec:G-fs-register}). This can be rewritten to make fewer calls.
    \item We currently replay all non-blocking collective communications, like \texttt{MPI\_Ibarrier},  \texttt{MPI\_Ireduce} and  \texttt{MPI\_Ibcast}, in order to re-create virtualized requests. Not only is time wasted by creating completed requests, but this also increases the size of the virtual request table and slows the translation between real requests and virtual requests.
\end{enumerate}

\subsection{Stragglers}
\label{sec:J-stragglers}

A \emph{straggler} is an MPI process participating in a collective communication that may take minutes to hours to join the collective communication, because it is finishing a CPU-intensive operation.  As a result, the completion of the collective communication is delayed.  Even worse, from the viewpoint of checkpointing, no checkpoint can take place while some processes are still in the middle of a collective call in the lower-half MPI library.

The original algorithm used a two-phase commit algorithm for collective communication, in order to be able to transparently checkpoint without significant waits (see~\cite{garg2019mana}).  That algorithm inserted a barrier operation before every collective communication call, and was found to result in runtime slowdowns.  A modified algorithm was introduced in a revised implementation.  The modified algorithm assumed that there were no stragglers, but that modified algorithm was found to have some flaws.  The current MANA-2.0 now includes a hybrid algorithm that adds a barrier operation before collective communication calls only after the DMTCP coordinator has requested a checkpoint.  Additional details outside the scope of this paper are included in the MANA-2.0 documentation~\cite{mana-documentation}.

\subsection{Globally unique IDs: \texttt{MPI\_Translate\_group\_ranks}}
\label{sec:K-globallyUniqueID}

A challenge in the previous implementation of the two-phase-commit algorithm is that the MANA centralized coordinator does not know which MPI processes participate in a given active communicator.  This limits the ability of the MANA centralized coordinator to determine which MPI processes must stop and await the final checkpoint command, and which MPI processes must continue to execute in order to ``unblock'' later collective communication calls.

In order to get around this issue in MANA-2.0, each MPI process reports to the centralized coordinator whether it is currently executing within a collective communication call --- and if so, provides a globally unique ID for that collective communicator.  For performance reasons, the process must compute this globally unique ID \emph{without the overhead of additional communication with its peers}.  This is done using \texttt{MPI\_Translate\_group\_ranks}.  This function enables the MPI process to translate the ranks of the current communicator to the corresponding rank in \texttt{MPI\_COMM\_WORLD}.  The ranks in the current communicator are all known as 0, 1, $\ldots$, \texttt{MPI\_Comm\_size()}-1, where we have taken liberties with the syntax of \texttt{MPI\_Comm\_size()}.   \texttt{MPI\_Translate\_group\_ranks} then produces the set of corresponding ranks in \texttt{MPI\_COMM\_WORLD}, and a hash function is used to produce an integer that is globally unique with high probability.

\subsection{Hybrid two-phase-commit algorithm (out of scope for this article)}
\label{sec:L-hybrid}

To further improve the performance of the two-phase-commit algorithm, we have designed a hybrid version of the algorithm that removes the barrier before each collective communication.  Additional details of this improved two-phase-commit algorithm are outside the scope of this paper; they are available in the MANA-2.0 documentation~\cite{mana-documentation}.

\subsection{Lessons learned}
\label{sec:M-lessonsLearned}
There are some lessons learned from building the algorithms discussed above. First, additional communication by MANA should be minimized to optimize performance and correctness. When possible, MANA-2.0 uses MPI calls for internally sharing information among processes, instead of relying on MANA's centralized coordinator. Also when possible, MPI calls that complete based on local information are preferred over MPI calls requiring peer-to-peer communication. Minimizing the internal communication of MANA not only improves runtime performance, but also reduces the possibility of race conditions in the code, and helps to simplify debugging.

Another lesson learned in MANA-2.0 is that some MPI calls can be emulated with other MPI calls. This was described in greater detail in item~1 at the beginning of Section~\ref{sec:componentAlgorithms}.

Last but not least, one should instrument MANA to provide additional information that can be used in its algorithms:  a globally unique ID for each communicator (see \texttt{MPI\_Translate\_group\_ranks}); and recording the number of bytes sent and received for each possible sender-receiver pair (using  \texttt{MPI\_Alltoall}). 

\section{Experimental Evaluation}
\label{sec:experiment}
All experiments were run on Cori, a Cray XC40 system at NERSC. Cori contains two types of compute nodes, dual-socket Intel Haswell and single-socket KNL nodes,  interconnected with Cray Aries network.
Each Haswell node has 32 cores (64 hardware threads) running at 2.3 GHz, and 128 GB DDR4 2133 MHz memory; each KNL node has 68 cores (272 hardware threads) running at 1.4 GHz, and 96 GB DDR4 2400 MHz memory. Cori runs Cray Linux environment version 7.0.UP01 with Linux kernel version 4.12.
All experiments used Cori's burst buffer~\cite{cori-bb} for I/O, the most suitable file system for writing checkpoint images on Cori. 


We evaluated MANA-2.0 using two commonly used applications at NERSC: GROMACS, a molecular dynamics code, and VASP, a materials science code. GROMACS (2021.02) was compiled with the Intel compiler (2019.3.199), and linked with Cray MPICH 7.7.10 and FFTW 3.3.8.  Two versions of VASP were tested: VASP 5 (5.4.4), a pure MPI code and VASP 6 (6.2.1), a hybrid OpenMP + MPI code. Both VASP versions were compiled with the Intel compiler (2019.3.199), and linked with Cray MPICH (7.7.10), MKL (2019.3.199) and FFTW (3.3.4) libraries. 

Two branches of MANA-2.0 were used in the experiments: 
a relatively stable branch, the master branch, and a development branch, ``feature/2pc''. The ``feature/2pc'' branch has been extensively tested (and is hence more reliable), but has a high runtime overhead. We used the master branch in the checkpoint/restart experiments. The ``feature/2pc'' branch contains the latest improvements to runtime overhead, thus we used it in the runtime overhead tests. There are some issues to resolve in interface8 before merging it into the stable master branch, e.g., it fails to restart GROMACS using 2048 MPI processes.
MANA is free and open-source software~\cite{mana-software}.  Documentation of the internals of MANA can be found at~\cite{mana-documentation}.

\subsection{Running GROMACS at Scale}

GROMACS was chosen to evaluate the scalability improvement of MANA after the code enhancements described earlier. 
GROMACS was run with MANA-2.0 on a AuCoo monolayer system containing 407,156 atoms (nano particles in water). This system was studied in~\cite{giorgia2021} by a NERSC user. 

First, we evaluated the runtime overhead of MANA by running the benchmark using from 1 to 64 Haswell and KNL nodes (strong scaling) with and without MANA. We used the interface8 branch of MANA with commits present at the time of this writing.  The interface8 branch includes mainly runtime overhead performance improvements.

\begin{figure}[!tb]
\centering
\includegraphics[width=0.8\columnwidth]{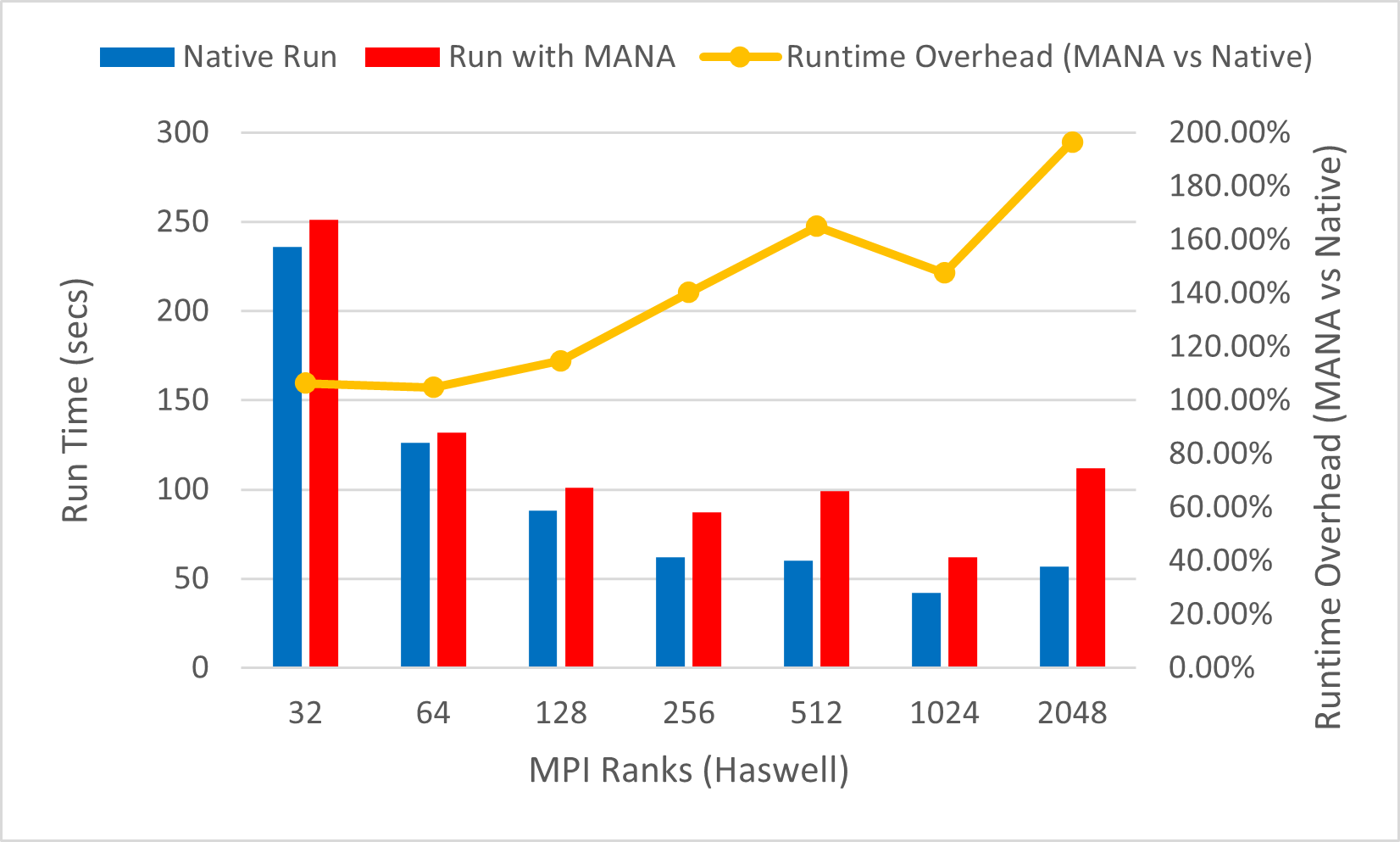}
\includegraphics[width=0.8\columnwidth]{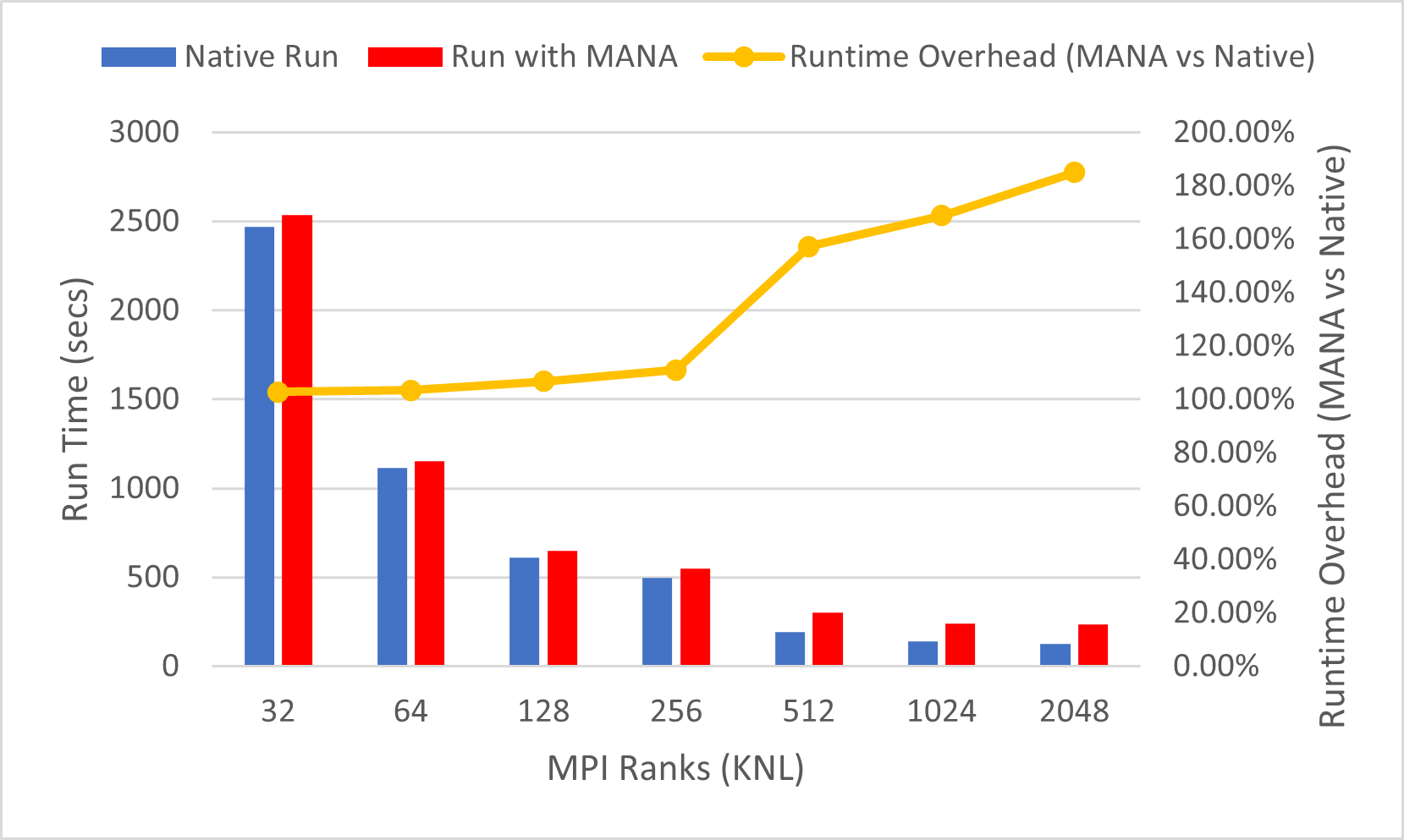}
\caption{Run time comparison between running GROMACS natively (blue bars) and under MANA (red bars) on Haswell (upper panel) and KNL (lower panel) nodes. The yellow line represents the run-time ratio between the MANA-enabled and native runs. Experiments were run on Cori Haswell and KNL nodes using 32 MPI processes per node. For the KNL runs, each task was run with two OpenMP threads.}
\label{fig:gromacs_overhead}
\end{figure}

The GROMACS run time was measured for 10,000 MD steps.
Figure~\ref{fig:gromacs_overhead} shows the results. The blue and red bars show the run time of GROMACS when running natively and under MANA, respectively; the yellow line shows the run-time ratio between the MANA-enabled and native runs. One can see that for KNL runs, the runtime overhead is negligible except at 2048 processes.  But for the Haswell architecture, the runtime overhead is still excessively high when running on more than two nodes, and increases rapidly when the number of processes increases. We continue our efforts to further reduce MANA's runtime overhead by addressing the remaining causes of the overhead. 

Next, MANA's checkpoint/restart capability was tested at scale.  GROMACS was run with MANA using 2048 processes using 64 Haswell and KNL nodes (32~processes per node), respectively.  The KNL runs were configured to use two OpenMP threads per process. The jobs were checkpointed at the 5-minute mark and terminated after 8 minutes (to assure sufficient time to write the checkpoint file), and then restarted. MANA was able to successfully checkpoint and restart GROMACS 10~times on each of Haswell and KNL.

Note that GROMACS does not scale well to an increasing number of MPI processes (see Figure~\ref{fig:gromacs_overhead}).  So using 2048 MPI processes to run the selected benchmark is not optimal. In fact, a high load imbalance was observed when running with 2048~MPI processes.
This, however, does not affect our goal of demonstrating the scalability of MANA. 

\begin{figure}[!tb]
\centering
\includegraphics[width=0.8\columnwidth]{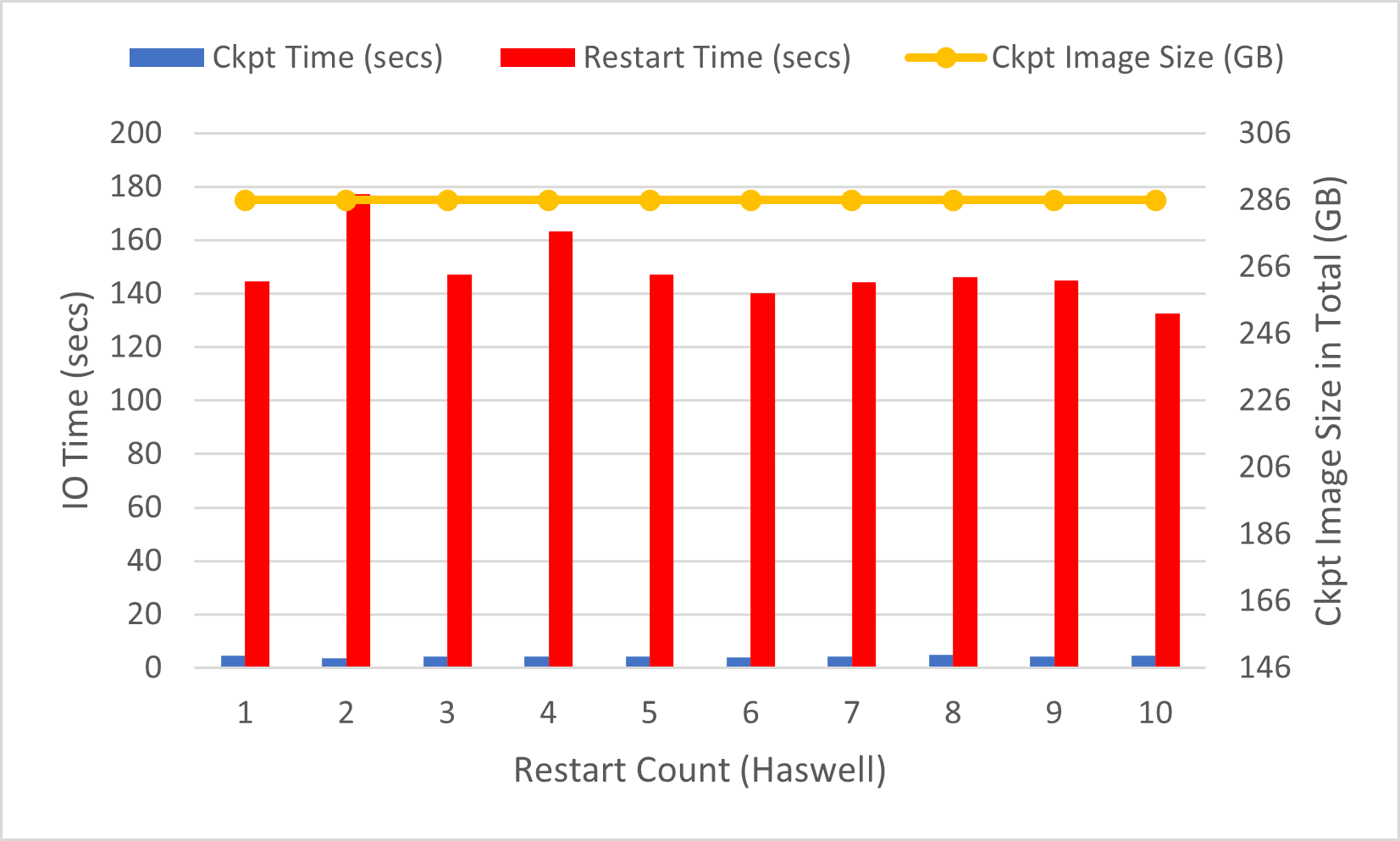}
\includegraphics[width=0.8\columnwidth]{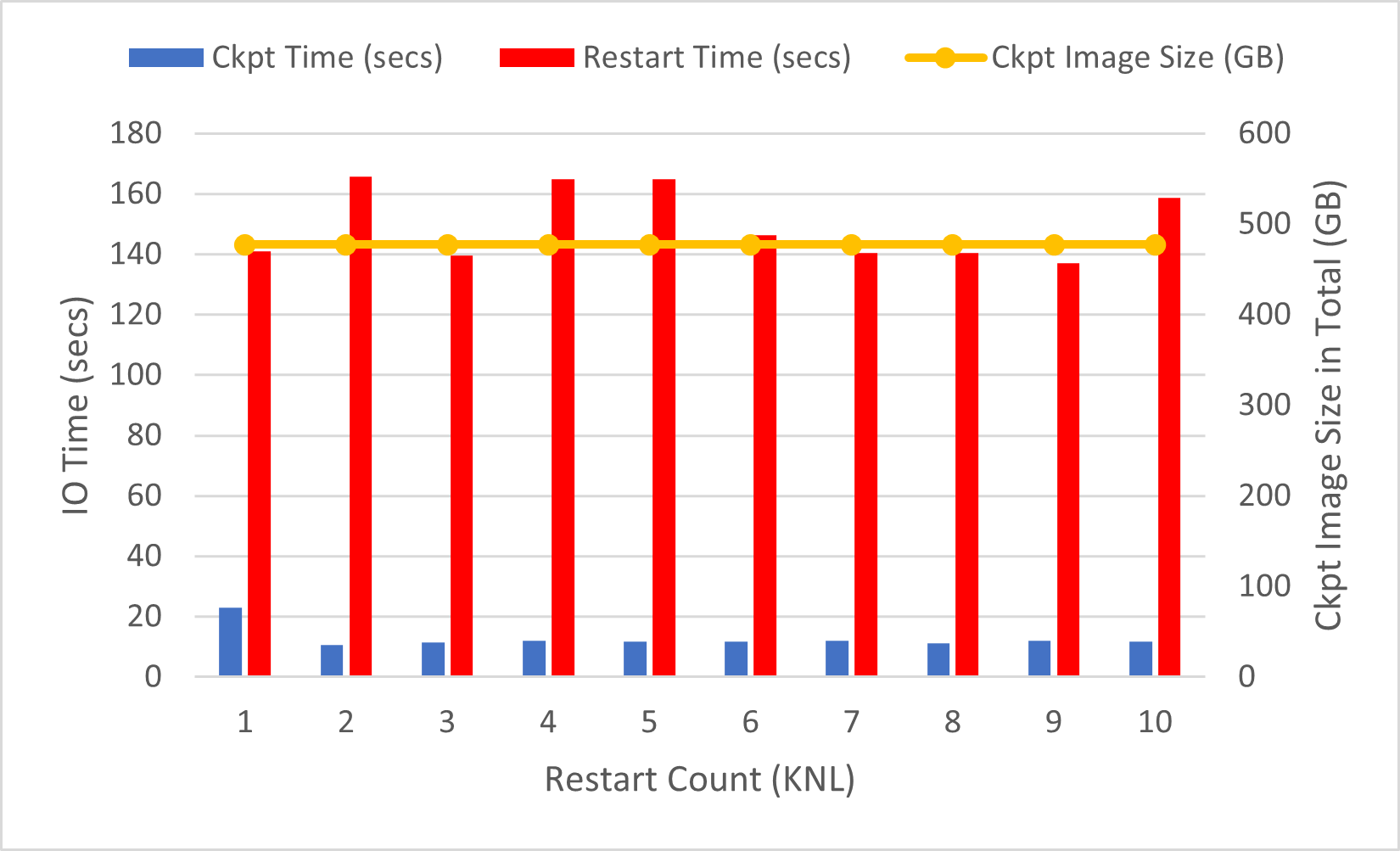}
\caption{Checkpoint/Restart overhead of MANA when running GROMACS with 2048 processes on Haswell (upper panel) and KNL (lower panel) nodes on Cori's Burst Buffer. The blue and red bars show the checkpoint and restart time, respectively; the yellow line indicates the total size of the checkpoint files.}
\label{fig:gromacs_cr_overhead}
\end{figure}



\subsection{VASP:  a resource for robustness testing}


\setlength{\tabcolsep}{3pt}
\begin{table*}[!htbp]
\center
\caption{\label{tab:app__info} VASP test cases for MANA-2.0. These cases were chosen to cover representative workloads and to exercise different code paths.}
\begin{tabular}{|p{0.75in}||p{0.55in}|p{0.55in}|p{0.55in}|p{0.55in}|p{0.6in}|p{0.55in}|p{0.55in}|p{0.55in}|p{0.6in}|}
\hline
  {}	& {\bf PdO4} & {\bf GaAsBi-64} & {\bf CuC\_vdw} & {\bf Si256\_hse} & {\bf B.hR105\_hse}  & {\bf PdO2} & {\bf CaPOH} & {\bf WOSiH}&{\bf GaAs-GW0}\\
\hline\hline
{\bf Electrons (Ions)} & 3288 (348) &266 (64)  &1064 (98)  &1020 (255) &315 (105)   &1644 (174)&288 (44)  &80 (18)& 8(2)\\
\hline
{\bf Functional} &DFT  &DFT  &VDW  &HSE & HSE &DFT &DFT & HSE&GW0\\
\hline
{\bf Algo} &RMM \newline (VeryFast) 	& BD+RMM \newline (Fast)	& RMM \newline (VeryFast)	&CG \newline (Damped) 	&CG \newline (Damped) 	&RMM \newline (VeryFast) &BD \newline (Normal)  & BD+RMM \newline (Fast) &BD \newline (Normal)\\
\hline
{\bf KPOINTS} &1 1 1  	& 4 4 4	& 3 3 1 & 1 1 1		&  1 1 1& 1 1 1 & 2 1 1 & 3 3 3& 3 3 3 \\
\hline
\end{tabular}
\end{table*}

\setlength{\tabcolsep}{3pt}
\begin{table*}[!htbp]
\center
\caption{\label{tab:interface8} Performance comparison of the VASP CaPOH workload with 128 ranks.}
\begin{tabular}{|p{0.6in}||p{0.6in}|p{0.9in}|p{0.9in}|}
\hline
  {}	& {\bf Native} & {\bf MANA \newline master branch} & {\bf MANA feature/2pc branch}\\
\hline\hline
{\bf Haswell} & 25s & 41s & 35s \\
\hline
{\bf KNL} & 69s & 137s & 101s \\
\hline
\end{tabular}
\end{table*}

\begin{figure}[!tb]
\centering
\includegraphics[width=0.8\columnwidth]{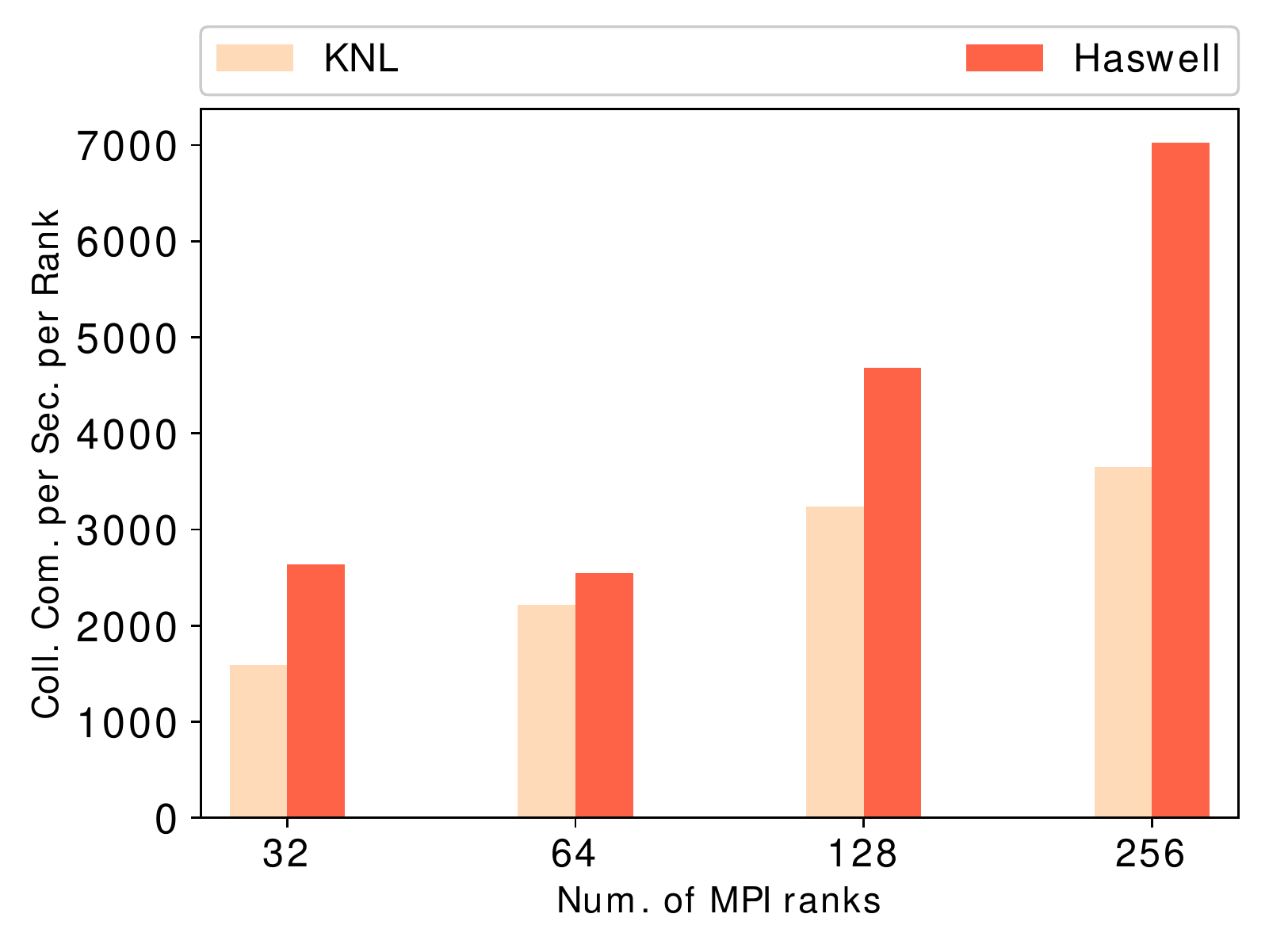}
\caption{Number of collective communications per second per process for VASP-5 on Haswell and KNL nodes. When doubling the number of ranks, the growth in the number of collective calls is roughly logarithmic
in the number of nodes. This figure was cited from~\cite{yao2021removing}.}
\label{fig:vasp_collectives}
\end{figure}

MANA-2.0 was tested with VASP, a materials science code.  VASP is largest consumer of computer time on Cori at NERSC. VASP has been extensively tested with MANA using the representative workloads summarized in Table~\ref{tab:app__info}. These benchmark cases were chosen to cover the representative VASP workloads and to exercise different code paths. For example, the first test case, denoted as PdO4 in the table, is a PdO slab containing 348~atoms. It was chosen to test the most commonly used code path, the DFT (Density Functional Theory) functional calculations using the RMM-DIIS~\cite{VASP-scheme} iteration scheme. Some test cases are real-world experiments, taken from the work of NERSC users.  

Many of the MANA code enhancements described earlier arose from fixing bugs and issues exposed by these VASP jobs when running them in production settings. As of this writing MANA-2.0 can successfully checkpoint and restart all the benchmark cases listed in Table~\ref{tab:app__info} with both VASP 5 (MPI) and VASP 6 (OpenMP + OpenMP). For VASP 6 we needed to disable the use of \texttt{MPI\_Win\_} family APIs at compilation time to use MANA, because they are not yet supported in MANA. There are still other issues to resolve, e.g., some of the VASP jobs run into segmentation faults after many rounds of checkpoint/restart.     

Note that VASP is highlighted for its intensive use of MPI collective communication.  While users typically run VASP across a small number of  nodes
due to the nature of its algorithms, VASP invokes an excessively high number of MPI collectives per second, as shown in Figure~\ref{fig:vasp_collectives}.  This presents an additional challenge:  runtime overhead. 

\subsection{MANA ``Hybrid-2PC'': initial successes in reducing runtime overhead}


VASP was chosen to evaluate improvement in runtime overhead. We tested the CaPOH workload with 128~processes on both Haswell and KNL nodes. Table~\ref{tab:interface8} shows the performances of the native VASP program, and VASP run under the MANA master branch and MANA experimental ``feature/2pc'' branch.

The MANA master branch focuses on scalability and stability; the ``feature/2pc" branch is our first attempt to reduce runtime overhead. Currently, ``feature/2pc" includes the hybrid two-phase-commit algorithm and removes lambda functions in the code base. From the table, we can see that on Haswell nodes, the runtime overhead has been reduced from 64\% to 40\%. On KNL nodes, the runtime overhead has been reduced from 99\% to 46\%. As discussed in Section~\ref{sec:componentAlgorithms}, there are additional known sources of runtime overhead. MANA-2.0 remains a work in progress, and efforts to solve each problem and reduce MANA's runtime overhead are continuing.

\section{Related Work}
\label{sec:relatedWork}

The history of MPI checkpointing is littered with approaches that tried too closely to tie the checkpointing process to a specific underlying network.

There have also been a series of checkpointing approaches for
particular implementations of MPI.  These include:  the Open~MPI
checkpoint-restart service~\cite{hursey2007design,hursey2009interconnect},
the MVAPICH2 checkpoint-restart service~\cite{gao2006application} ---
both of which temporarily disconnect the network and then delegate
to BLCR~\cite{hargrove2006blcr} for checkpointing an individual process.
Similarly, MPICH-V~\cite{bouteiller2006mpich} disconnects a transport layer channel of MPICH (primarily based on TCP).
It then delegates to the Condor package for checkpointing single-threaded individual processes~\cite{litzkow1997checkpoint}.

Each of the above packages is implemented within a particular implementation of MPI.  In contrast,
the original DMTCP~\cite{ansel2009dmtcp} (for TCP), and an InfiniBand plugin for DMTCP~\cite{cao2014transparent} are independent of the MPI implementation, and do not disconnect the network during checkpoint.
But both are otherwise tightly bound to the underlying network.

An approach to support mobile MPI applications exists, albeit while partially abandoning application transparency and requiring re-compilation of the MPI application source code~\cite{fernandes2006mobile}.
And CIFTS provides a fault-tolerant BLCR-based ``backplane''~\cite{gupta2009cifts}.

MANA then introduced the \emph{split-process model} for checkpointing of MPI~\cite{garg2019mana}.
Details are in Section~\ref{sec:splitProcess}.  
The first work~\cite{garg2019mana} demonstrated transparent checkpointing of GROMACS~\cite{gromacs} and additional three applications at 2048 MPI processes over 64 Haswell nodes.
Efforts to deploy MANA at NERSC
are described in~\cite{zhao2020deploying}, while MANA was previously
updated for compatibility with the latest NERSC environment and to remove
code specific to one environment,
as described in~\cite{chouhan2021improving}.
The second work~\cite{chouhan2021improving} demonstrated transparent checkpointing for 64 processes with GROMACS and 512 processes with the HPCG benchmark~\cite{dongarra2016new}.

Note that transparent checkpointing of MPI was already demonstrated to the level of 16,368 processes for NAMD and 32,368 processes for HPCG (using 1/3 of the supercomputer) on Stampede at TACC in 2016~\cite{cao2016system}.
That early work was based on DMTCP's transparent support for InfiniBand, and that code likely would not run on today's machines using either Cray GNI or extensions to the original InfiniBand.
Further, no effort was made in the current work to test the limits of scalability of MANA-2.0.

\section{Conclusion}
\label{sec:conclusion}

This report on MANA-2.0 represents encouraging progress toward a robust, reliable package for transparent checkpointing that will be future-proof.  
There are important lessons from this work.  
Each individual subsystem for MANA-2.0 must be carefully designed with appropriate data structures and algorithms to enable an MPI computation to survive over the checkpoint-restart barrier.  The subsystems requiring particular support are:  point-to-point communication (translating \texttt{MPI\_Send} to \texttt{MPI\_Isend}, etc.); MPI collective communication (allowing each MPI process to proceed until all MPI processes have reached a safe point with no MPI process currently in an MPI call);  decisions whether to wait for an MPI call to complete, or to virtualize and replay at restart time;  MPI requests (virtualizing those requests and deciding when the memory of old requests can be reclaimed); and in the case of asynchronous MPI calls, deciding which processes must replay point-to-point and collective calls in order to re-instantiate vitual MPI requests for completion after restart.

MANA-2.0 is a significant improvement over the previous MANA in both scalability and reliability. It can checkpoint and restart GROMACS reliably, which uses MPI point-to-point communication intensively, using up to 2048 MPI processes (with the selected benchmark system). It can also checkpoint and restart the representative production workloads of VASP, which uses MPI collective communications intensively. 

MANA-2.0 is still a work in progress. There are multiple issues to resolve before it can be used in production. Currently applications that invoke MPI collectives frequently, or applications that run at large scale, incur high runtime and memory overheads. Many causes have been identified, and fixes have been implemented or are under implementation.

In particular, MANA-2.0 need not be restricted to a standalone environment.  It would be simple to extend MANA-2.0 to support the MPI-3.1 tools interfaces.  This would give MANA-2.0 the ability to introspect into an MPI implementation.  Hence, MANA-2.0 could play a supportive role within other fault-tolerant libraries.  And the use of the tools interface could lighten MANA's current burden of indirect discovery through wrapper functions around MPI calls.  The tools interface also represents an opportunity to provide a deadlock detector, as one more component in a general fault-tolerant ecosphere that includes such well-known packages as CRAFT~\cite{shahzad2018craft} (Checkpoint/Restart and Automatic Fault Tolerance), SCR~\cite{moody2010design} (Scalable Checkpoint/Restart library), ULFM~\cite{laguna2016evaluating} (User-Level Failure Mitigation), and VeloC~\cite{nicolae2019veloc} (Very Low Overhead
Checkpoint-Restart).




\section*{Acknowledgments}

The authors thank two NERSC users, Giorgia Brancolini (Institute of Nanoscience CNR--Modena) and Daniel Pert (University of Michigan), for providing the benchmark cases for GROMACS and VASP (WOSiH), respectively, and for providing detailed descriptions of the systems. The authors also thank Chris Samuel at NERSC for valuable discussions and help, and Prashant Chouhan for his insights into the issues of allowing MANA to run at production scale. The authors wish to thank the reviewers for their insightful comments.  They would especially like to thank the particular reviewer who provided many detailed suggestions for the future improvement of MANA-2.0.
This  work  used  the  resources  of  the National  Energy  Scientific Computing Center (NERSC) at the Lawrence Berkeley National Laboratory.

\IEEEtriggeratref{28}  
\bibliographystyle{IEEEtran}
\bibliography{supercheck21-sc}

\end{document}